\documentclass[aps,prx,twocolumn,groupedaddress]{revtex4-1}
\usepackage[english]{babel}
\usepackage[percent]{overpic}

\usepackage{verbatim}
\usepackage[utf8]{inputenc}
\usepackage{amsmath,amssymb,amsfonts}
\usepackage{color}
\usepackage{hyperref}
\usepackage{filecontents}
\usepackage{graphicx}
\usepackage{soul}
\usepackage{bm}
\usepackage{makeidx}
\usepackage{wasysym}
\makeindex
\def\>{\right\rangle}
\def\<{\left\langle}
\def\be{\begin{equation}}
\def\ee{\end{equation}}
\def\ba{\begin{array}{l}}
\def\ea{\end{array}}

\def\beq{\begin{eqnarray}}
\def\eeq{\end{eqnarray}}

\begin{document}
\title{Interference induced thermoelectric switching and heat rectification in quantum Hall junctions}
\author{Luca Vannucci$^{1,2}$, Flavio Ronetti$^1$, Giacomo Dolcetto$^2$, Matteo Carrega$^2$, Maura Sassetti$^{1,2}$}
\affiliation{ $^1$ Dipartimento di Fisica, Universit\`a di Genova,Via Dodecaneso 33, 16146, Genova, Italy.\\
$^2$ CNR-SPIN, Via Dodecaneso 33, 16146, Genova, Italy.
} 
\date{\today}
\begin{abstract}
Interference represents one of the most striking manifestation of quantum physics in low-dimensional systems. Despite evidences of quantum interference in charge transport have been known for a long time, only recently signatures of interference induced thermal properties have been reported, paving the way for the phase-coherent manipulation of heat in mesoscopic devices.
In this work we show that anomalous thermoelectric properties and efficient heat rectification can be achieved by exploiting the phase-coherent edge states of quantum Hall systems.
By considering a tunneling geometry with multiple quantum point contacts, we demonstrate that the interference paths effectively break the electron-hole symmetry, allowing for a thermoelectric charge current flowing either from hot to cold or viceversa, depending on the details of the tunnel junction.
Correspondingly, an interference induced heat current is predicted, and we are able to explain these results in terms of an intuitive physical picture.
Moreover, we show that heat rectification can be achieved by coupling two quantum Hall systems with different filling factors, and that this effect can be enhanced by exploiting the interference properties of the tunnel junction.
\end{abstract}
\maketitle
\section{Introduction}
\label{sect:intro}
In recent years a great attention has been devoted to the study of thermal transport at the nanoscale and energy and heat exchanges in small quantum devices~\cite{esposito2009, campisi2011, giazottorev, lirev, segal13, saito2007, lopez14, sothmann15, carrega15}. A deep understanding of these phenomena is of paramount importance for applications in solid-state cooling, high-precision sensors, cryogenic thermometry and thermal logic in quantum information \cite{giazottorev, lirev, dubi, benentirev}. This emerging field goes under the name of caloritronics \cite{giazottorev, giazotto12, martinez-perez14a, martinez-perez14b}. Here, surprising experimental results have been already obtained. Among them Giazotto {\it et al.} have recently demonstrated the possibility to coherently manipulate the heat flux in a hybrid superconducting circuit \cite{giazotto12, martinez-perez14a, martinez-perez14b}.
By realizing the thermal version of the electric Josephson interferometer, they have paved the way towards the phase-coherent manipulation of heat in mesoscopic devices \cite{simmonds12}.
Moreover, implementations of mesoscopic heat engines and thermal diodes have been proposed and will be soon achieved \cite{giazottorev, fornieri14, matthews14, ruokola09, segal05}.
These can be realized using quantum dots \cite{sothmann15, hwang14, d-sanchez13, ruokola11}, optomechanical systems \cite{mari2014}, and multiterminal mesoscopic devices combining also normal metals, superconductors and ferromagnets \cite{r-sanchez15, bosisio15, mazza15}.
The rapid progresses made in the field of caloritronics allow to glimpse a future in which electric and thermal manipulation will proceed on equal footing.\\
In this context a promising role is played by topologically protected states, such as the edge states of quantum Hall systems (QHS) and topological insulators (TI).
As far as the recently discovered TIs are concerned, the presence of protected helical edge states not only allows to generate peculiar spin-dependent thermal phenomena \cite{arrachea11, rothe12}, but also to achieve high thermoelectric performances \cite{takahashi10, takahashi12, yokoyama14, xu14}.
On the other hand, the interest of the scientific community for QHS has been refueled in the view of possible thermal applications\cite{lopez14, r-sanchez15, grosfeld, fradkin13}.
For example, Sanchez {\it et al.} have demonstrated that a three terminal device in the quantum Hall regime can work as a perfect thermal diode, with a rectification coefficient $r_Q \to \infty$, exploiting the chirality of edge states \cite{r-sanchez15}. Heat transport measurements were also proposed in order to extract important information on fractional statistics and neutral modes in exotic fractional quantum Hall states\cite{grosfeld, fradkin13}.
In these topologically states protection from backscattering guarantees phase-coherent ballistic transport of charge and heat over long ($\apprge$ $\mu$m) distances \cite{konig07, roth09, olshanetsky15, kononov15, eisenstein09}, and the possibility to control their charge transport properties in the presence of applied bias voltages via interferometric setups have been extensively studied. However, the effects of quantum interference on their thermal and thermoelectric properties remain quite unexplored, and our work, motivated indeed by the recent interest in phase-coherent heat manipulation, moves in this direction.
   
   In this paper we show that the chiral edge states of QHS can be exploited to implement coherent caloritronics.
In particular, we consider two separated QHS coupled by a tunneling region, driven out of equilibrium by the presence of a thermal gradient that induces finite charge and heat flows.
We demonstrate that a tunnel junction with $n$ quantum point contacts (QPCs) enables to control the charge and heat transfer between the two QHS.
As far as the charge sector is concerned, we show that, by varying the geometrical parameters of the junction, particle-hole symmetry can be broken, allowing to selectively enhance electron tunneling with respect to hole tunneling or viceversa. Therefore we predict the interesting result that quantum interference can be exploited to selectively switch the charge flow induced by the thermal gradient, that is, charge current can either flows from hot to cold or from cold to hot, depending on the interference properties.\\
Remarkably, we find that quantum interference phenomena control heat transport as well.
Contrary to the charge transport, with electrons and holes giving opposite contributions, heat transport does not depend on the charge of the carriers, so that electrons and holes equally contribute; this property manifests in the correspondence between the zeros of the charge current and the maximum/minimum visibility of the heat current.
Furthermore, we study how these quantum interference phenomena, already present in the absence of interactions (tunneling between integer QHS), are affected by the presence of interactions, by considering tunneling of electrons between fractional and integer QHS, where e-e interactions play an important role.
As a general remark, a strong suppression of the signal appears, due to the anomalous temperature dependence of the effective tunneling density of states, a hallmark of non-Fermi liquid behavior.
More interestingly, the presence of different filling factors breaks the left-right symmetry, inducing rectification effects.
By taking advantage from the interference patterns induced in the presence of several QPCs, we show that strong rectification effects can be obtained. We thus demonstrate that the interplay between interactions and quantum interference is crucial in order to enhance the heat rectification.

The paper is organized as follows. In Sec.~\ref{sect:model} we describe the setup and evaluate the charge and heat currents for a generic tunneling region. Sec.~\ref{sect:discussion} is devoted to the main discussion, focusing on interference phenomena~(\ref{sect:interference}) and rectification effects~(\ref{sect:rectification}) in a multiple QPC geometry. Finally, we draw our conclusions in Sec.~\ref{sect:conclusion}.

\section{Model and transport equations}
\label{sect:model}
\begin{figure}[!ht]
	\includegraphics[width= 0.48\textwidth]{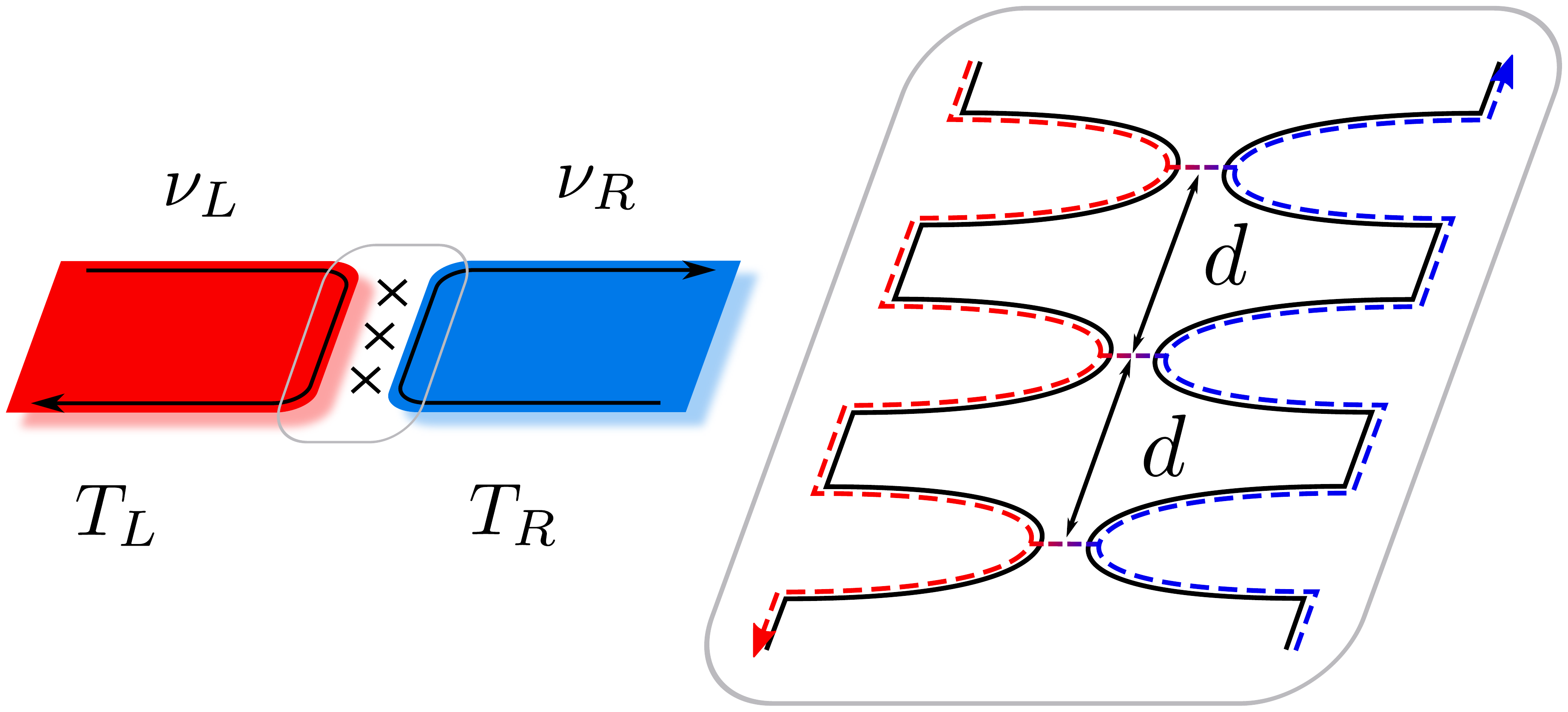}
\caption{(Color online) Scheme of two fractional quantum Hall systems  with different filling factors $\nu_L$ and $\nu_R$ at temperatures $T_L>T_R$. Counterpropagating edge states are coupled by a tunneling region, schematically depicted in the middle. The zoom in the right panel shows the case of tunneling due to multiple quantum point contacts ($n=3$ in this case) equally spaced with distance $d$.
}
\label{fig:one}
\end{figure}
We consider two quantum Hall bars with filling factor $\nu_\alpha$ ($\alpha = R,L$) belonging to the Laughlin sequence~\cite{laughlin83, tsui99}, with the same chemical potential $\mu=\mu_L=  v_L k_{{\rm F},L} =\mu_R= v_Rk_{{\rm F},R}$.
They are kept at two different temperatures $T_L>T_R$ and coupled by a tunneling region, as shown in Fig.~\ref{fig:one}. The two quantum Hall systems (QHS) have counterpropagating single edge channels with  Hamiltonian  (in this work $\hbar = k_{{\rm B}}=1$)
\be
\label{eq:h0}
H_\alpha = \frac{\pi v_\alpha}{\nu_\alpha} \int d x \rho_\alpha^2 (x)= \frac{v_\alpha}{4\pi \nu_\alpha}\int dx \left(\partial_x \phi_\alpha (x)\right)^2\,.
\ee
Here, $v_\alpha$ is the propagation velocity of the mode and $\nu_\alpha = 1/m_\alpha$, with $m_\alpha \geq 1$ an odd integer\cite{laughlin83}. The case  $m_\alpha=1$ corresponds to an integer QHS, while $m_\alpha >1$ describes fractional quantum Hall liquids. In the second expression of Eq.~(\ref{eq:h0}) the electron density $\rho_\alpha (x)$ is written in terms of the chiral bosonic particle-hole collective excitations field $\phi_\alpha(x)$. Using bosonization technique~\cite{giamarchi, miranda03} the electron operator $\psi_\alpha(x)$ can be also expressed in terms of  $\phi_\alpha(x)$, 
\be
\label{eq:psi}
\psi_\alpha(x) = \frac{{\cal F}_\alpha}{\sqrt{2\pi a}}e^{ i \alpha k_{{\rm F},\alpha} x} e^{ i\frac{\alpha}{\nu_\alpha}\phi_\alpha(x)}~,
\ee
with $a$ a short distance cut-off and ${\cal F}_\alpha$ the so-called Klein factor\cite{giamarchi}.
The index $\alpha = R (+ ) , L(-)$ indicates also the direction of propagation and $k_{{\rm F},\alpha}$ is the associated Fermi momentum. 
We assume that the two QHS are tunnel coupled with a tunneling Hamiltonian
\be
\label{eq:h_tun}
H_\Lambda = \Lambda \int d x f(x) \psi_R^\dagger (x) \psi_L(x) + {\rm H.C.}~,
\ee
where $f(x)$ describes the shape of the tunneling region~\cite{footnote-tunnel} and $\Lambda$ the amplitude strength \cite{chamon97, dolcetto12}. In the following we will consider a series of multiple  $n$ point-like contacts equally-spaced \cite{footnote-distance} with distance $d$~~\cite{chamon97,dolcini11,ferraro13, footnote-origin}, with $f(x)=\sum_{j=0}^{n-1} \delta (x - j d)/n$. By properly acting on the gate voltages of the QPCs one can manipulate and tune their transmissions, selectively opening or closing some of them.
The electric charge $J_C$ and heat $J_Q$ currents can be written  in terms of  particle and energy variations $J_N= \langle \dot{N}_R - \dot{N}_L\rangle /2$ and $J_H = \langle \dot{H}_R - \dot{H}_L\rangle /2$ as
\be
J_C = - e J_N\, \qquad J_Q = J_H - \mu J_N,
\ee
with $N_\alpha = \int dx \rho_\alpha(x)$ the particle numbers on each edge. The averages $\langle \dots \rangle $  are taken over the equilibrium states of the left and right QHS with respect to their temperatures $T_L$, $T_R$~\cite{footnote-phonons}. They can be computed  at lowest order in the tunneling, using standard perturbation techniques~\cite{martin05, kamenev09}
\beq
\label{eq:jc}
J_C &=& - 2 i e |\lambda|^2 \int dx  dx' \int d \tau f(x) f(x') \sin[2  \bar{k}_{{\rm F}}(x-x')] \nonumber \\
& \times &  P_{m_L}\left(\tau+\frac{x-x'}{v_L} \right) P_{m_R}\left(\tau - \frac{x-x'}{v_R} \right)~,
\eeq
and
\beq
\label{eq:jq}
J_Q &=& - i |\lambda|^2 \int dx  dx' \int d \tau f(x) f(x') \cos[2  \bar{k}_{{\rm F}}(x-x')] \nonumber \\
&\times & \left\{ \partial_\tau  P_{m_L}\left(\tau+\frac{x-x'}{v_L} \right) P_{m_R}\left(\tau - \frac{x-x'}{v_R} \right) \right. \nonumber \\
&& \left. - P_{m_L}\left(\tau+\frac{x-x'}{v_L} \right) \partial_\tau P_{m_R}\left(\tau - \frac{x-x'}{v_R} \right) \right\}.
\eeq
Here, $\lambda \equiv \Lambda/(2\pi a)$ and $\bar{k}_{{\rm F}}\equiv (k_{{\rm F}, R} + k_{{\rm F},L})/2$ the average Fermi momentum.
In Eqs.~(\ref{eq:jc})-(\ref{eq:jq}) we introduced the function $P_{m_\alpha}(t)=e^{m_\alpha {\cal W}_\alpha(t)}$, with ${\cal W}_\alpha(t)= \langle \phi_\alpha (t) \phi_\alpha(0)\rangle - \langle \phi_\alpha^2 (0)\rangle$
the bosonic correlator  given by\cite{giamarchi, vondelft98, ferraro14, dolcetto14}
\be
\label{eq:w}
{\cal W}_\alpha (t) =  \ln \frac{ \left| \Gamma \left(1+ \frac{T_\alpha}{\omega_{\rm c}} + i T_\alpha t \right) \right|^2}{\Gamma^2 \left(1+\frac{T_\alpha}{\omega_{\rm c}} \right)(1+i\omega_{\rm c} t)}~,
\ee
with $\Gamma(z)$ the Euler $\gamma$-function and $\omega_{{\rm c}}$ the high energy cut-off.
It is now useful to introduce the Fourier transform {$\hat{P}_{m_\alpha}(E) = \int dt e^{-i E t} P_{m_\alpha}(t)$}. In the energy representation we get
\beq
\label{eq:jc2}
J_C &=& - i e \frac{|\lambda|^2}{\pi} \int dx  dx' \int\limits_{-\infty}^{+\infty} dE f(x) f(x') \sin[2 \bar{k}_{{\rm F}}(x-x')] \nonumber \\
&\times &  e^{2i\frac{E}{\mu} \bar{k}_{{\rm F}} (x-x')} \hat P_{m_L}(E) \hat P_{m_R}(-E)\,, \\
J_Q &=& \frac{|\lambda|^2}{\pi} \int dx  dx' \int\limits_{-\infty}^{+\infty} dE f(x) f(x') \cos[2 \bar{k}_{{\rm F}}(x-x')] \nonumber \\
&\times &  e^{2i\frac{E}{\mu} \bar{k}_{{\rm F}} (x-x')} E \hat P_{m_L}(E) \hat P_{m_R}(-E)~.
\eeq
In the scaling limit $\omega_{{\rm c}}/T_\alpha \gg 1$, $\hat{P}_{m_\alpha}(E)$ can be conveniently recast as $\hat{P}_{m_\alpha}(E) ={\cal D}_{m_\alpha}(E) n_\alpha(E)$ with 
 \be
\label{eq:dos}
{\cal D}_{m_\alpha}(E) = \frac{(2\pi)^{m_\alpha}}{\omega_{\rm c} \Gamma(m_\alpha)} \left(\frac{T_\alpha}{\omega_{\rm c}}\right)^{m_\alpha-1} \frac{\left|\Gamma \left( \frac{m_\alpha}{2} + i \frac{E}{2 \pi T_\alpha}\right)\right|^2 }{\left|\Gamma \left( \frac{1}{2} + i \frac{E}{2 \pi T_\alpha}\right)\right|^{2} }~,
\ee
which plays the role of  an ``effective'' tunneling density of states (DOS)~\cite{sassetti94,furusaki02, vondelft98, kane} and $n_\alpha (E) = [e^{E/T_\alpha}+1]^{-1}$, the equilibrium Fermi distribution function at temperature $T_\alpha$~\cite{footnote-fermi}. Note that in the non-interacting/integer case ($\nu_\alpha=1$, $m_\alpha=1$) the DOS is constant, as for a normal Fermi liquid, while for the fractional case ${\cal D}_{m_\alpha}(E)$ is energy and temperature dependent taking into account the non-Fermi liquid nature of the fractional QHS\cite{sassetti94, furusaki02, vondelft98, kane}. Using the symmetry properties ${\cal D}_{m_\alpha}(E)={\cal D}_{m_\alpha}(-E)$, and  $n_\alpha (E) + n_\alpha (-E)=1$ the charge and heat currents assume the more compact form
\begin{equation}
	\label{eq:Landauer-Buttiker}
	\begin{split}
	\begin{pmatrix} J_C \\ J_Q \end{pmatrix} & = \frac{|\lambda|^2}{2\pi} \int\limits_{-\infty}^{+\infty}  d E \begin{pmatrix} -e \\  E \end{pmatrix} g(E+ \mu) \\
	& \times {\cal D}_{m_L}(E) {\cal D}_{m_R}(E) [n_L(E)-n_R(E)] ,
	\end{split}
\end{equation}
where we introduced the transmission function
\be
\label{eq:modulating}
g(E) = \int dx dx'f(x) f(x') \cos \left[\frac{2E}{\mu}\bar{k}_{{\rm F}} (x-x') \right]~.
\ee
In this way Eq.~(\ref{eq:Landauer-Buttiker}) takes an analogous form of the well-known Landauer-B\"uttiker expression for transport~\cite{sivan85,buttiker85,blanter00} with however a renormalized effective DOS ${\cal D}_{m_\alpha}(E)$.
The transmission is also sensitive to the shape of the tunneling constriction within the $g(E)$ function. 
For a periodic array of $n$ QPCs, the case of interest here, the modulating function is
\begin{equation}
	\label{eq:modulating_nQPC}
	g_n(E) = \frac{n + 2 \sum_{j=1}^{{n-1}} (n-j) \cos(2 j \eta E/\mu)}{n^2}~,
\end{equation}
with the dimensionless quantity $\eta= \bar{k}_{{\rm F}} d$. 

One immediately recognizes that $g_1(E) = 1$ in the case of single QPC, while oscillating functions of the form $\cos(2 j \eta E/\mu)$ appear for multiple QPCs $n\geq 2$.

\section{Interference induced thermoelectric phenomena}
\label{sect:discussion}
\subsection{Thermoelectric switching and heat current interference}
\label{sect:interference}
\begin{figure}[!ht]
	\includegraphics[width= 0.48\textwidth]{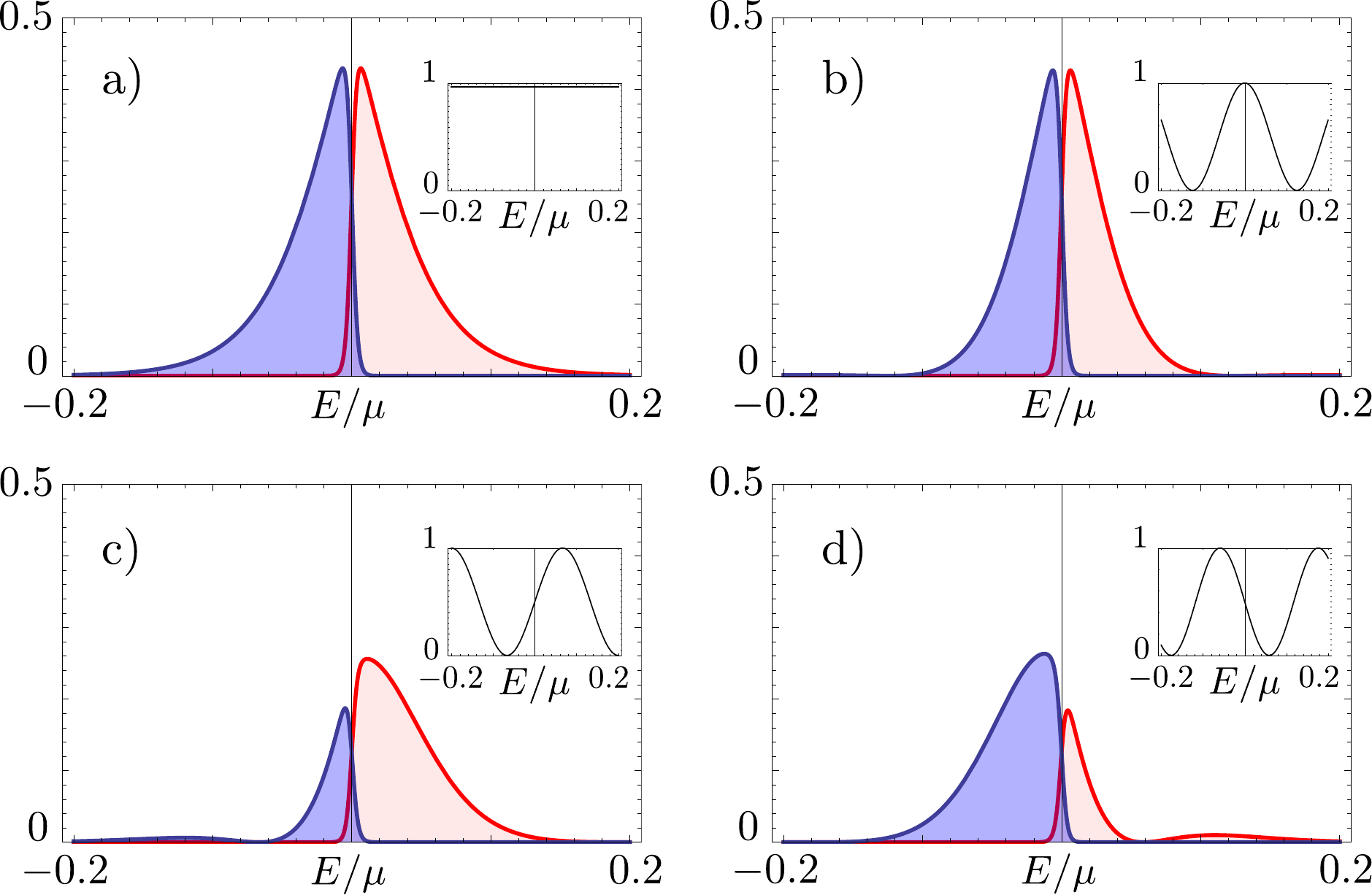}
\caption{(Color online) Schematic representation of electron and hole contributions to the charge current in the presence of a thermal gradient (here $T_L=300$~mK, $T_R=20$~mK, and $\mu=10$~K). The electron contribution $g_n(E+\mu)n_L(E)\left [1-n_R(E)\right ]$ is represented in red, while the hole one $g_n(E+\mu)\bar{n}_L(E)\left [1-\bar{n}_R(E)\right ]$ is represented in blue (see Eq. \eqref{eq:non-int2}). The insets show the transmission functions $g_n (E + \mu)$. (a) Single QPC: the transmission function is energy independent and electrons and holes contribute equally, thus giving $J_C=0$. (b-d) 2 QPCs. (b) $\eta=4\pi$: despite the transmission function is energy dependent, it does not break particle-hole symmetry, so that $J_C=0$. (c) $\eta=4\pi-\pi/4$: particle-hole symmetry is broken and electron tunneling is enhanced with respect to hole tunneling, giving $J_C<0$. (d) $\eta=4\pi+\pi/4$: particle-hole symmetry is broken and hole tunneling is enhanced with respect to electron tunneling, giving $J_C>0$. 
}
\label{fig:charge_scheme}
\end{figure}

We start the discussion with the two QHS at integer filling factors $\nu_L=\nu_R = 1$. Here, Eq.~(\ref{eq:Landauer-Buttiker}) reduces to

\begin{equation}
	\label{eq:non-int}
	\begin{pmatrix} J_C \\ J_Q \end{pmatrix} = \frac{2 \pi |\lambda|^2}{\omega_{\rm c}^2} \int\limits_{-\infty}^{+\infty} dE\begin{pmatrix} -e \\ E \end{pmatrix} g_n(E +\mu) [n_L(E)-n_R(E)] .
\end{equation}
Explicit calculation, inserting Eq.~(\ref{eq:modulating_nQPC}), leads to
\be
\label{eq:jc_nu1}
{J}_C = \frac{4\pi |\lambda|^2 e\mu }{\omega_{{\rm c}}^2}\sum_{j=1}^{n-1}2(n-j)\frac{\sin(2j\eta)}{2j\eta n^2} {\cal I}_3\left(\frac{2j\eta T_L}{\mu},\frac{2j\eta T_R}{\mu}\right)
\ee
and
\beq
\label{eq:jq_nu1}
{J}_Q&=& \frac{4\pi |\lambda |^2 \mu^2}{\omega^2_{{\rm c}}}\left\{\frac{1}{n}{\cal I}_1\left(\frac{T_L}{\mu},\frac{T_R}{\mu}\right)\right.\nonumber \\
&+&\sum_{j=1}^{n-1} 2(n-j)\frac{\cos(2j\eta)}{(2j\eta)^2 n^2}{\cal I}_2\left.\left(\frac{2j\eta T_L}{\mu},\frac{2j\eta T_R}{\mu}\right)\!\!\right\}
\eeq
where
\begin{align}
{\cal I}_1(x,y) & = \frac{\pi^2}{12}(x^2-y^2)\\
	{\cal I}_2(x,y) & = \frac{\pi^2}{2} \left[\frac{x^2 \cosh(\pi x)}{\sinh^2(\pi x)}-\frac{y^2 \cosh(\pi y)}{\sinh^2(\pi y)} \right]\\
	{\cal I}_3(x,y) & = \frac{\pi}{2} \left[\frac{x}{\sinh(\pi x)}-\frac{y}{\sinh(\pi y)} \right]~.
\label{functions}
\end{align}
Before discussing the results, in order to make realistic predictions, it is useful to restrict the parameter range to a set of experimentally reasonable one. We thus estimate $\bar{k}_{{\rm F}}\sim 1/\ell_B$ with $\ell_B \sim 10$~nm  a typical magnetic length of QHS\cite{altimiras, chevallier10}. The velocity of the edge states $v_\alpha$ is of the order of $\sim 10^4$m$/$s. Using these values we have $\mu \sim 10$~K.
We set the temperature range between $20-300$~mK, typical values in which well-developed fractional quantum Hall plateau were measured\cite{tsui99, altimiras}.
Furthermore, we consider distances between the contacts of the order $10-300$~nm (less than the phase coherence length at the considered temperatures~\cite{eisenstein09}), corresponding to a dimensionless parameter range $\eta \sim 1 - 30$.
With these parameters the QPCs separation $d$ is never much larger than the thermal lengths $L_{\alpha}=v_{\alpha}/T_{\alpha}$\cite{lehur02, virtanen11}.\\
\begin{figure}
	\begin{overpic}[width=\linewidth]{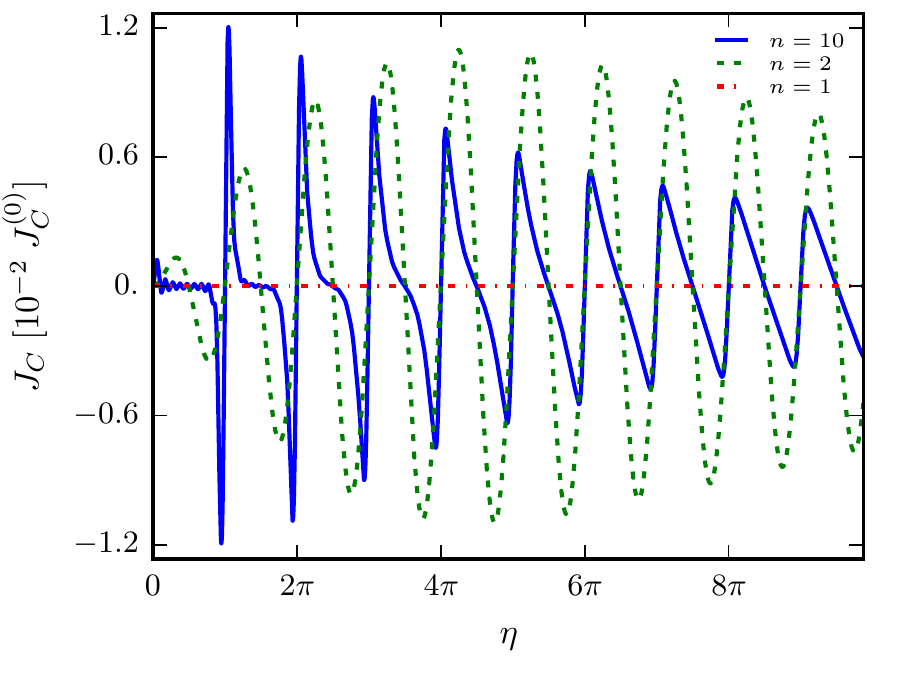}\put(1,70){a)}\end{overpic}
	\begin{overpic}[width=\linewidth]{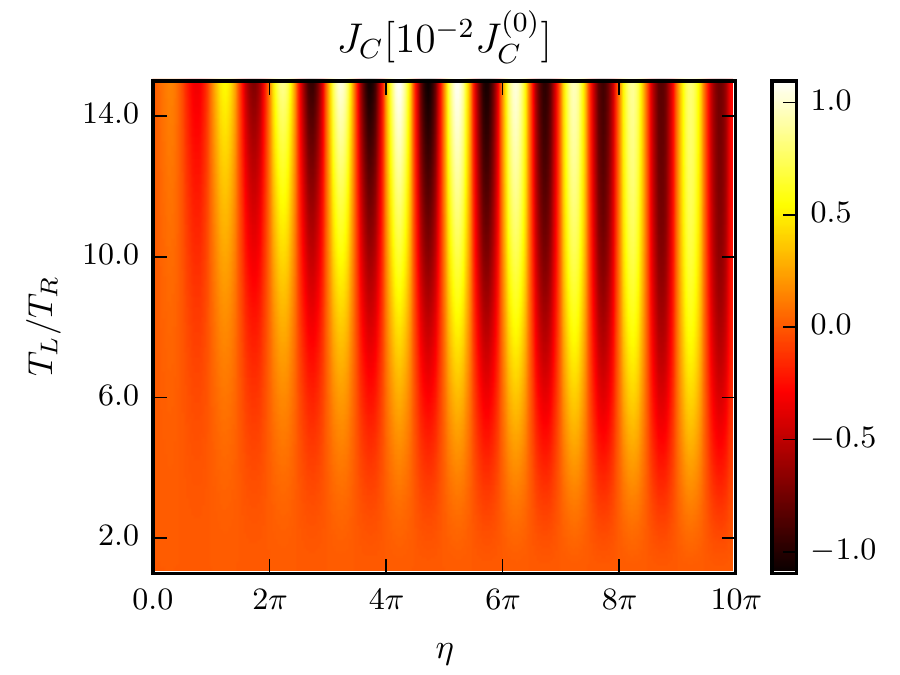}\put(1,70){b)}\end{overpic}
	\caption{(Color online) Tunneling charge current $J_C$ in units of $J_C^{(0)}= 2 \pi |\lambda|^2 e \mu /\omega^2_{{\rm c}}$ for $\nu_L=\nu_R=1$, as a function of $\eta = \bar{k}_{{\rm F}} d$. Panel a): different curves represent different numbers of QPCs with temperature $T_R=20$~mK and $T_L=300$~mK. The $n=2$ curve displays oscillations of period $\pi$, modulated by a non-monotonic envelope function that reaches its maximum for $\eta \sim 4.5\pi$. More complicated interference patterns arise for the $n=10$ curve. Note that the single QPC ($n=1$) doesn't break particle-hole symmetry, resulting in a null charge current. Panel b): Density plot of charge current for $n=2$ QPCs as a function of $\eta$ ($x$ axis) and $T_L/T_R$ ($y$ axis). $T_R$ is fixed and equal to $20$~mK, while the ratio $T_L/T_R$ goes from 1 to 15 (corresponding to a maximum temperature of $300$~mK). The figure shows an increasing or decreasing monotonic behavior of the charge current as a function of the ratio $T_L/T_R$ at fixed $\eta$, depending on the sign of $J_C$. The oscillating behavior described in panel a) is also visible.
In both panels the chemical potential is set to $\mu=10$ K.
}
	\label{fig:two}
\end{figure}
We start now to analyze the charge current. Fig.~\ref{fig:two}a) shows $J_C$ as a function of $\eta$ for different number of QPCs at fixed $T_L>T_R$.
In the case of a single QPC $n=1$ (red/dashed-dotted curve), the charge current is always zero, because the energy independence of $g_1(E)=1$, see Eq.~\eqref{eq:modulating_nQPC}, does not induce particle-hole symmetry breaking~\cite{kane}, so that electrons and holes equally contribute to transport leading to no net charge current. On the other hand, in the presence of a multiple QPC setup ($n=2$, $n=10$ in the Figure) quantum interference phenomena are responsible for an energy-dependent transmission function $g_n(E)$, effectively breaking in general particle-hole symmetry and leading to a non-vanishing charge current. Interestingly enough, the charge current exhibits an oscillating  behavior, switching between positive and negative values, with principal zeros at $\eta = k \pi/2$ ($k$ integer). 
This result suggests that, despite the thermal gradient is fixed, charge can flow either from hot to cold or from cold to hot, depending on the parameters of the junction only.
To shed light on this result, it is useful to rewrite Eq.~\eqref{eq:non-int} as
\begin{eqnarray}
	\label{eq:non-int2}
	J_C &=& \frac{2 \pi |\lambda|^2}{\omega_{\rm c}^2} \left \{-e\int\limits_{-\infty}^{+\infty} dE g_n(E +\mu)n_L(E)\left [1-n_R(E)\right ]\right . \nonumber\\
&+&\left .e \int\limits_{-\infty}^{+\infty} dE g_n(E +\mu)\bar{n}_L(E)\left [1-\bar{n}_R(E)\right ]\right \}
,\end{eqnarray}
with $\bar{n}_{\alpha}(E)=n_{\alpha}(-E)$ representing the Fermi distribution for holes and where the first (second) line represents the electron (hole) contribution to the charge transport from left to right.
In this picture, transport is due either to electron tunneling, \textit{i.e.} $n_L(E)\left [1-n_R(E)\right ]\neq 0$, or to hole tunneling \textit{i.e.} $\bar{n}_L(E)\left [1-\bar{n}_R(E)\right ]\neq 0$. Obviously, these two cases differ for the sign of the carriers, as shown in Eq.~\eqref{eq:non-int2}, so that if particle-hole symmetry is present, no charge current is expected.
In order for the charge current to be finite, the transmission function must break the particle-hole symmetry, differently weighting electron and hole contributions. This cannot happen for a single QPC with $g_1(E)=1$, as schematically shown in Fig. \ref{fig:charge_scheme}a). However, if $n>1$ QPCs are present, the energy dependent transmission function can promote electron tunneling with respect to hole tunneling or viceversa, inducing either negative or positive charge current respectively.
In particular, transitions between positive and negative values of the charge current occur for $\eta = k \pi /2$, where $g_n(E+\mu)=g_n(-E+\mu)$: at these specific values particle-hole symmetry is restored, {\it i.e.} electrons and holes contribute equally giving a null charge current signal, as shown in Fig. \ref{fig:charge_scheme}b).
Quite interestingly the sign of the current switches between positive and negative values despite the presence of a fixed thermal gradient direction $T_L>T_R$. This simply reflects a change in the majority of the carriers: either electrons, Fig. \ref{fig:charge_scheme}c), or holes, Fig. \ref{fig:charge_scheme}d). This argument remains valid also for $n> 2$, but the presence of higher harmonics shifts the position of maximal intensity, as shown in Fig. \ref{fig:two}a).

\noindent Acting on the parameter $\eta$ one can therefore switch the sign (and thus the flow) of the charge current: this can be used to implement a device that, exploiting quantum interference, allows to selectively switch the charge flow induced by a fixed thermal bias. To complete the description, we present in Fig.~\ref{fig:two}b) the density plot of the charge current for $n=2$ as a function of the temperature ratio $T_L/T_R$ and $\eta$.
The switching behavior of the charge current is stable against temperature variations. The oscillations as a function of $\eta$ have the same zeros also with varying temperature. 
The interference patterns are modulated by an envelope function which moves towards higher $\eta$ values while lowering the temperature ratio. They show a power law behavior $(T_L/T_R)^{2}$ varying the thermal gradient and as a function of $\eta$ a  dephasing envelope which scales as $1/\eta$ for large $\eta$ values,  (see Eq.~(\ref{eq:jc_nu1})). Note that the crossing to an exponential dephasing dependence would be present only at much larger temperatures/QPCs separation, out of the considered parameter range~\cite{virtanen11}.\\ 
\begin{figure}
	\begin{overpic}[width=\linewidth]{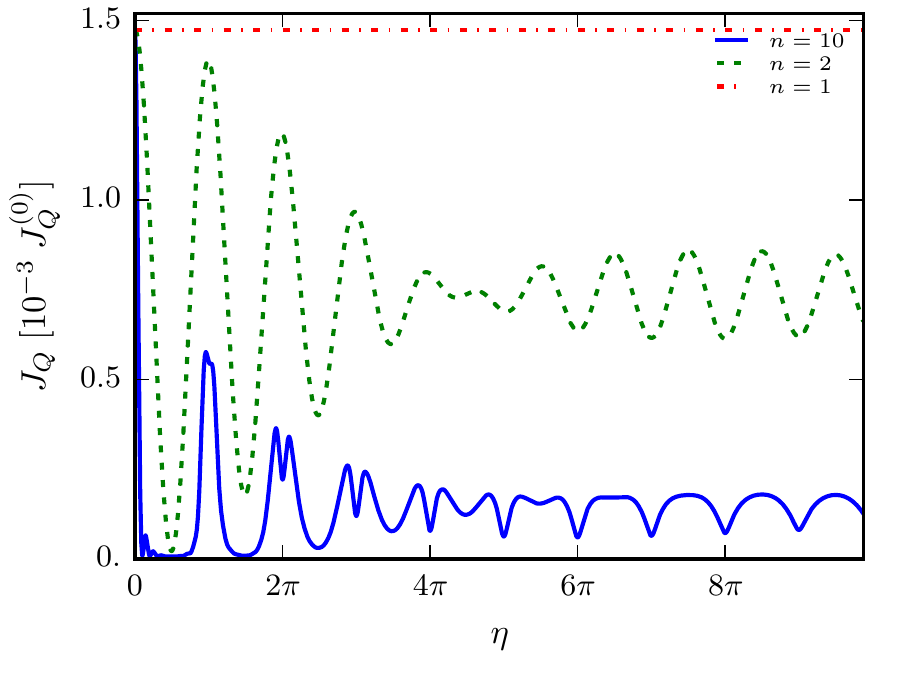}\put(1,70){a)}\end{overpic}
	\begin{overpic}[width=\linewidth]{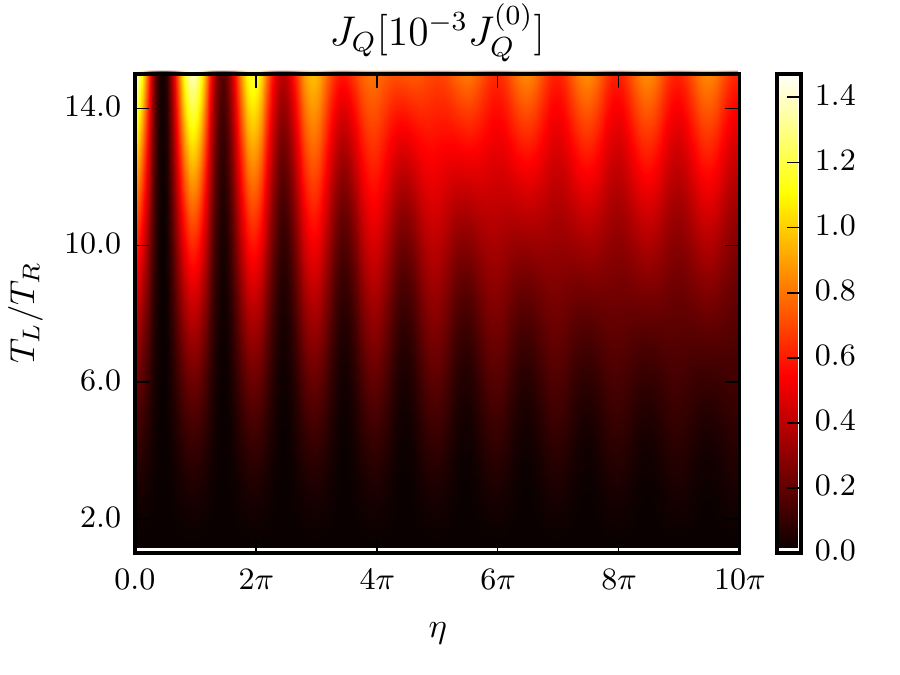}\put(1,70){b)}\end{overpic}
	\caption{(Color online) Tunneling heat current $J_Q$ in units of $J_Q^{(0)}= 2 \pi |\lambda|^2  \mu^2 /\omega^2_{{\rm c}}$ for $\nu_L=\nu_R=1$. The chemical potential is set to $\mu=10$~K.
	Panel a): $J_Q$ as a function of $\eta = \bar{k}_{{\rm F}} d$.  different curves represent different number of QPCs with temperature $T_R=20$~mK and $T_L=300$~mK. As for the charge current, oscillations of period $\pi$ are visible in the $n=2$ signal. Here heat current oscillates around a non vanishing mean value equals to half the value of the single QPC heat current. Note that the modulating function of the oscillating part of the signal changes its sign at $\eta \sim 4.5\pi$, displaying a sort of beat. More complicated interference patterns arises for the $n=10$ curve. Panel b): density plot of heat current for $n=2$ QPCs as a function of $\eta$ ($x$ axis) and $T_L/T_R$ ($y$ axis). $T_R$ is fixed and equal to $20$~mK, while the ratio $T_L/T_R$ goes from 1 to 15 (corresponding to a maximum temperature of $300$~mK). The heat current is an increasing monotonic function of the ratio $T_L/T_R$ for all values of $\eta$. The oscillating behavior described in panel a) is also visible.
	}
	\label{fig:three}
\end{figure}
The arguments exposed above explain also the oscillating interference patterns of the heat current ${J}_Q$ shown in Fig.~\ref{fig:three}.
Indeed, as done for the charge current, one can rewrite the heat current as
\begin{eqnarray}
	\label{eq:non-int_h}
	J_Q &=& \frac{2 \pi |\lambda|^2}{\omega_{\rm c}^2}\left \{\int\limits_{-\infty}^{+\infty} dE E g_n(E +\mu)n_L(E)\left [1-n_R(E)\right ] \right .\nonumber\\
&+&\left .\int\limits_{-\infty}^{+\infty} dE (-E) g_n(E +\mu)\bar{n}_L(E)\left [1-\bar{n}_R(E)\right ]\right \}
.\end{eqnarray}
Contrary to Eq. \eqref{eq:non-int2}, electron and hole contributions add up, because the heat current is insensitive to the charge of the carriers.
The main features are represented by the presence of minima and maxima as a function of $\eta$, around the mean value proportional to  ${\cal I}_1$ in Eq.~(\ref{eq:jq_nu1}). For $n=2$ they coincide with $\eta = (2 k + 1)\pi/2$ and $\eta = k \pi$ respectively, and correspond to values at which the transmission function at zero energy has a minimum or a maximum respectively. These are precisely the values that give zero charge current. Then the heat current has a maximum or a minimum if, due to quantum interference, electrons and holes have high or low transmission respectively, as schematically represented in Fig. \ref{fig:heat_scheme}.
\begin{figure}[!ht]
	\includegraphics[width= 0.48\textwidth]{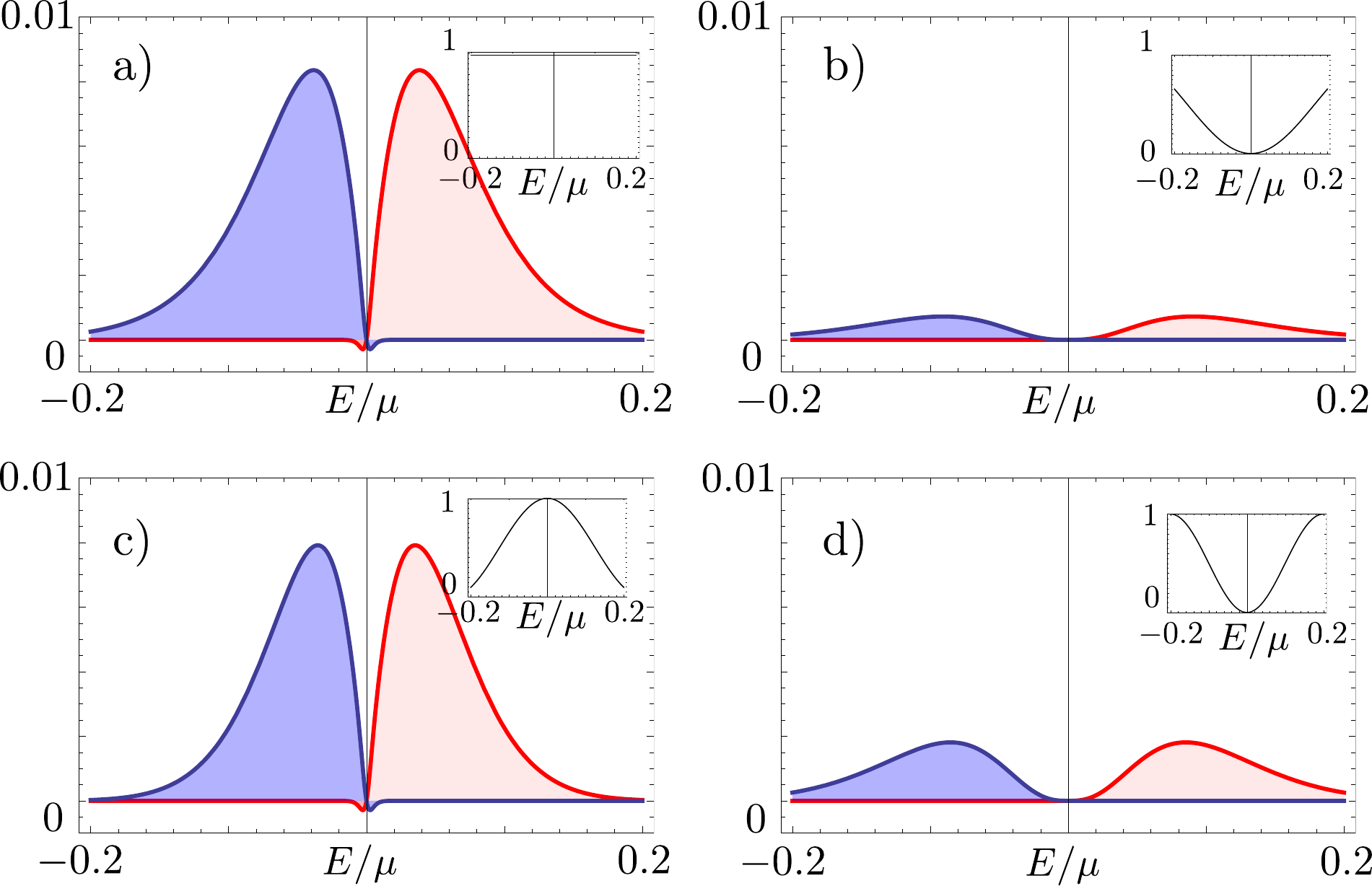}
\caption{(Color online) Schematic representation of electron and hole contributions to the heat current (in units of $\mu$) in the presence of a thermal gradient (here $T_L=300$ mK, $T_R=20$ mK, and $\mu=10$~K). The electron contribution $Eg_n(E+\mu)n_L(E)\left [1-n_R(E)\right ]$ is represented in red, while the hole one $(-E)g_n(E+\mu)\bar{n}_L(E)\left [1-\bar{n}_R(E)\right ]$ is represented in blue (see Eq. \eqref{eq:non-int_h}). Contrary to the charge current, the total heat current is given by the sum (and not the difference) of electron and hole contributions, see Eq. \eqref{eq:non-int_h}. The insets show the transmission functions $g_n(E+\mu)$. (a) Single QPC: the transmission function $g_1(E)=1$, giving the maximum heat current, see Fig. \ref{fig:three}a). (b-d) 2 QPCs. The heat current shows a sequence of maxima and minima (compare with Fig. \ref{fig:three}). (b) $\eta=2\pi-\pi/2$: both electrons and holes have low transmission, giving a minimum of $J_Q$. (c) $\eta=2\pi$: both electrons and holes have high transmission, giving a maximum of $J_Q$. (d) $\eta=2\pi+\pi/2$: both electrons and holes have low transmission, giving a minimum of $J_Q$. }
\label{fig:heat_scheme}
\end{figure}
Generalizing to $n$ contacts, there are multiple paths whose phase differences are always multiples of $2\eta$. This explains the more complicated interference patterns shown in Fig. \ref{fig:three}a) for $n=10$.
Similar curves are obtained for different values of temperatures as shown in the density plot of Fig.~\ref{fig:three} b) for $n=2$. Here, the magnitude of the interference patterns increase  with temperature following the power law $(T_L/T_R)^2$ increasing the temperature gradient.  
Note that in both panels at $\eta$ bigger than a critical value  ${\eta}_c$ the interference paths have a phase shift of $\pi$, with $\eta_c$ becoming lower and lower by increasing the number of QPCs.\\
{
Mathematically, this is due to a change in the sign of the envelope function ${\cal I}_2$ in Eq.~(\ref{functions}).
On a more physical ground, one can observe that states contributing to heat transport are mostly distributed around $E\sim\pm \bar{T}=\pm (T_L + T_R)/2$, as can be argued from Fig. \ref{fig:heat_scheme}(a).
Therefore, the behavior of the heat current depends on whether these states have high or low transmission, that is, if $\left .g_n(E+\mu)\right |_{E\sim \bar{T}}\approx(\ll) 1$ maxima (minima) of the heat current are expected. 
}
We focus on the simple case $n=2$, where the transmission function shows an oscillating pattern with period $\Delta E=\pi\mu/\eta$.
Then, consider what happens for values of $\eta$ multiple of $\pi$, corresponding to maximum transmission at zero energy, \textit{i.e.} $\left .g_2(E+\mu)\right |_{E=0}=1$.
In this case, as long as $\Delta E\gg 4\pi \bar{T}$ the transmission function is slowly oscillating so that states contributing to heat transport have high transmission, since $\left .g_2(E+\mu)\right |_{E\sim \bar{T}}\approx 1$, and a maximum of the heat current appears.
On the other hand, if $\Delta E\ll 4\pi \bar{T}$ the transmission function rapidly oscillates, so that $\left .g_n(E+\mu)\right |_{E\sim \bar{T}}$ can be significantly smaller than one, leading to a minimum of the  current.
This mechanism is related to the presence of thermal dephasing and induces an exchange between maxima and minima by increasing $\Delta E$, that is, by increasing $\eta$, as shown in Fig.~\ref{fig:three} a).
One can roughly estimate the crossover as $\Delta E_{{\rm c}} \approx 4\pi \bar{T}$, which gives $\eta_{{\rm c}}\approx \frac{\mu}{4\bar{T}}$, and corresponds indeed to a critical length $d_{{\rm c}}\approx L_T$, with $L_T=\frac{\mu/\bar{k}_{\mathrm{F}}}{\bar{T}}$ the characteristic thermal length.
Note that in the case of $n>2$ QPCs the presence of additional modulations in the transmission function leads to a decrease of the critical value $\eta_{{\rm c}}$.

\begin{figure}
	\begin{overpic}[width=\linewidth]{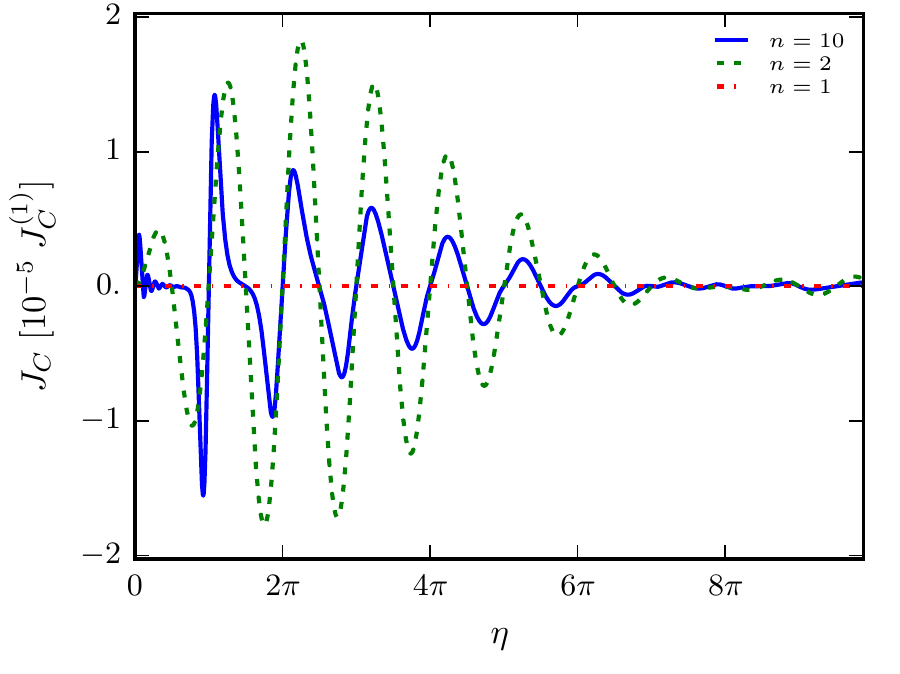}\put(1,70){a)}\end{overpic}
	\begin{overpic}[width=\linewidth]{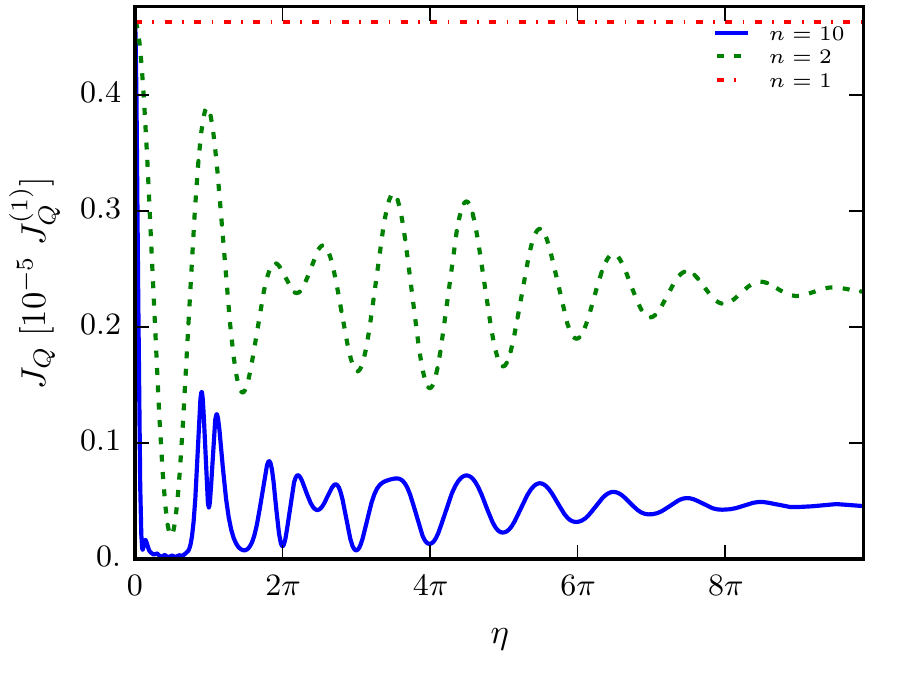}\put(1,70){b)}\end{overpic}
	\caption{(Color online) Charge (panel a) and heat (panel b) currents for $\nu_L=1$ and $\nu_R=1/3$ as a function of $\eta=\bar{k}_{{\rm F}} d$ at fixed temperatures $T_R=20$~mK and $T_L=300$~mK. Units of $J_C$ or $J_Q$ are respectively $J_C^{(1)} = 2\pi |\lambda|^2 e \mu^3/\omega^4_{{\rm c}}$and  $J_Q^{(1)} = 2\pi |\lambda|^2 \mu^4/\omega^4_{{\rm c}}$. Different curvers refer to different number of QPCs. The chemical potential is set to $\mu=10$ K.
}
	\label{fig:four}
\end{figure}
We now comment on the tunneling between different filling factors $\nu_R \neq \nu_L$ with fractional QHS. For sake of simplicity, we focus on the two bars with $\nu_L=1$ and $\nu_R=1/3$. Generalization to other filling factors~\cite{tsui99, ferraro14} are straightforward.
Also in this case one can calculate Eq.~(\ref{eq:Landauer-Buttiker}) in analytical form (not quoted). Fig.~\ref{fig:four}a) shows the charge current $J_C$ and Fig.~\ref{fig:four}b) shows the heat current $J_Q$ as a function of $\eta$ at fixed $T_L/T_R=15$ for different number of QPCs. The oscillating behaviors are again present, reflecting the interference patterns. The charge current has the same positions of the zeros of the integer case.  However, the currents show a  faster decreasing of the visibility, while increasing $\eta$.  
This is an hallmark of the fractional nature reflected in the peculiar behavior of the effective  DOS ${\cal D}_{m_R}(E)$. Indeed, the latter acquires an energy and temperature dependence, with the well-known power-law dependence of a non-Fermi liquid, thus modifying the envelope function. Here also the scaling behavior with temperature is modified, which is reflected in a strong suppression of the signal for both $J_C$ and $J_Q$.

\subsection{Heat current rectification}
\label{sect:rectification}
We now focus on thermal rectification effects.
We consider the heat current $J_Q$ in two different (and opposite) configurations: forward $J_Q^{(f)}\equiv J_Q(T_L=T_{{\rm hot}}, T_R=T_{{\rm cold}})$ and backward $J_Q^{(b)}\equiv J_Q(T_L=T_{{\rm cold}}, T_R=T_{{\rm hot}})$, exchanging the temperatures $T_{{\rm hot}}$ and $T_{{\rm cold}}$.    
One can conveniently define a rectification coefficient as\cite{fornieri14, martinez-perez15}
\be
\label{eq:r}
r_Q = \left| \frac{J_Q^{(f)}}{J_Q^{(b)}}~\right| .
\ee
Recalling Eq.~(\ref{eq:Landauer-Buttiker}), for constant DOS ${\cal D}_{m_R}(E)$ and ${\cal D}_{m_L}(E)$ (integer/non interacting case) one obtains $r_Q=1$, where the difference between forward and backward heat currents is a simple change of the sign.
We now show that rectification effects with $r_Q \neq 1$ are possible in the case of tunneling between a fractional QHS and an integer one, achieving also large rectification coefficients. This phenomenon is related to the energy and temperature dependent effective DOS ${\cal D}_{m_\alpha}(E)$ proper of a fractional filling. In this case, the difference in the DOS between $\nu_R$ and $\nu_L$ breaks the left-right symmetry, see Eq.~(\ref{eq:Landauer-Buttiker}), allowing for $r_Q \neq 1$.
For sake of simplicity, hereafter we restrict the discussion to the tunneling between $\nu_R=1/3$ and $\nu_L=1$. However, provided that $\nu_R \neq \nu_L$, all results and conclusions remain valid and can be easily generalized to other filling factors belonging to the Laughlin sequence. We point out that for larger differences $m_R - m_L$ one would obtain stronger rectification effects ({\it e.g.} $\nu_R=1/5$-$\nu_L=1$ gives higher rectification coefficient compared to the case $\nu_R=1/3$-$\nu_L=1$).

\begin{figure}
	\begin{overpic}[width=\linewidth]{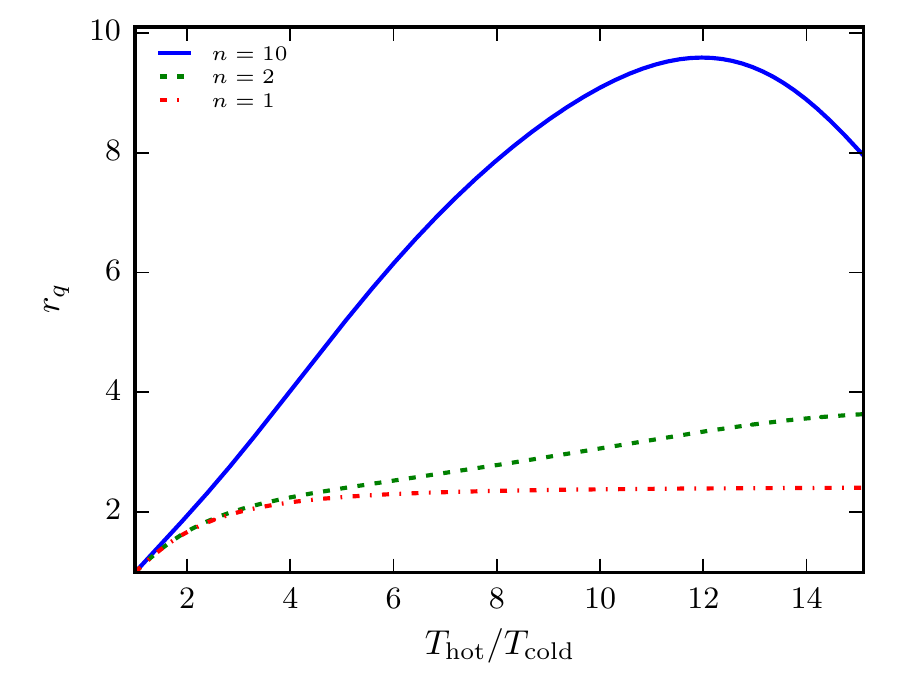}\put(1,70){a)}\end{overpic}
	\begin{overpic}[width=\linewidth]{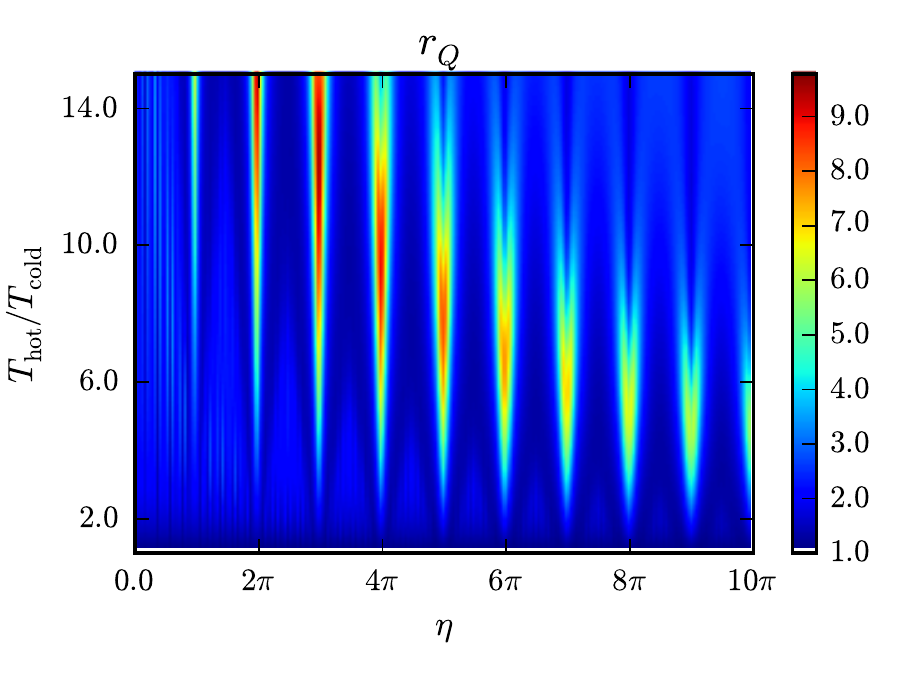}\put(1,70){b)}\end{overpic}
	\caption{(Color online) Panel a): heat rectification coefficient $r_Q$ for multiple QPC geometry as a function of $T_{{\rm hot}}/T_{{\rm cold}}$ for $\eta=3\pi$. Different curves refer to different number of QPCs. In the case of single QPC (solid line) $r_Q$ saturates to $2.43$ as one can see from Eq.~(\ref{eq:rq_1qpc}). Panel b): density plot of the heat rectification coefficient as a function of $\eta$ ($x$ axis) and $T_{{\rm hot}}/T_{{\rm cold}}$ ($y$ axis) for $n=10$ QPCs.
 The ratio $T_{{\rm hot}}/T_{{\rm cold}}$ goes from 1 to 15, with fixed $T_{{\rm cold}}=20$~mK. Larger rectification effect occurs when $\eta = k\pi$ and is linked to interference patterns. 
 For each value of $k$ the rectification coefficient is a non-monotonic function of $T_{{\rm hot}}/T_{{\rm cold}}$, displaying a maximum whose position changes with $k$. For $\eta = 3 \pi$ and $T_{{\rm hot}}/T_{{\rm cold}} \sim 12$ one can reach a maximum value of $r_Q \sim 9.6$.
}
	\label{fig:five}
\end{figure}

In the case of single QPC it is possible to get a simple expression for the rectification coefficient
\be
\label{eq:rq_1qpc}
r_Q=\frac{ 17 \tau^4 - 10 \tau^2 - 7}{7\tau^4 + 10 \tau^2 - 17}~
,\ee
in terms of $\tau = T_{{\rm hot}}/T_{{\rm cold}}$.
Note that for $T_{{\rm hot}}=T_{{\rm cold}}$ one has $r_Q=1$ as expected.
On the other hand, in the limit $T_{{\rm hot}} \gg T_{{\rm cold}}$ ($\tau \gg 1$) the rectification coefficient saturates to the value $r_Q \to 17/7 =2.43$.\\
Moreover, it is interesting to study possible enhancements of the rectification effects due to quantum interference phenomena arising in the presence of $n$ multiple QPCs, which, as we have shown in the previous Section, have a strong impact on heat transport.
In Fig.~\ref{fig:five}a) we show the rectification coefficient as a function of $T_{{\rm hot}}/T_{{\rm cold}}$ at fixed $T_{{\rm cold}}=20$~mK, for different number of QPCs.
The case of single contact (solid line in the figure) is an increasing function of $\tau$ as reported in Eq.~(\ref{eq:rq_1qpc}).
The different qualitative behavior of $r_Q$ in the presence of $n>1$ QPCs is due to the interplay between interactions and interference effects.
Remarkably, a pronounced peak can be observed, which drifts to lower temperatures, with higher values, while increasing the number of contacts $n$.
In Fig.~\ref{fig:five}b) we report a density plot of $r_Q$ as a function of $\eta=\bar{k}_{{\rm F}} d$ ($x$ axis) and $T_{{\rm hot}}/T_{{\rm cold}}$ ($y$ axis).
Interestingly, the optimal condition for the enhancement of $r_Q$ corresponds to values of $\eta$ for which no charge current flows in the system.\\
We have shown that efficient heat rectification performances can be achieved by increasing the number of QPCs. There are however limitations, related to the requirement that phase coherence is preserved throughout the tunneling paths, thus giving rise to quantum interference phenomena. This constrain limits the number of QPCs that can be created to $n<l_{\mathrm{in}}/d$, with $l_{\mathrm{in}}$ the inelastic mean free path, that can be of the order of several $\mu$m in QHS.

\section{Conclusions}
\label{sect:conclusion}

We studied charge and heat transport in two temperature-biased QHS coupled by a tunneling region. We showed that, when the tunneling is realized via a series of point-like contacts, an interference mechanism take place for both charge and heat currents. A multiple QPCs geometry can effectively break particle-hole symmetry, leading to a finite thermoelectric charge current.
Remarkably its sign, that is, the charge current flowing either from hot to cold or viceversa, is governed by quantum interference and can be manipulated.
Interference effects affect thermal transport as well, with the heat current displaying peculiar oscillations as a function of the distance between the QPCs.
We explained both the heat current oscillations and the thermoelectric switching in terms of different transmission functions for electrons and holes, due to the different tunneling paths.
Moreover, heat rectification can be achieved when considering two fractional QHS with different filling factors, due to the anomalous non-Fermi liquid tunneling density of states of the Laughlin state.
Despite heat flow is rectified already in the single contact geometry, the presence of multiple QPCs allows to exploit the interferometric properties to find an optimal working condition for the enhancement of the heat rectification effects.
\section{Acknowledgments}
We acknowledge the support of the MIUR-FIRB2012 - Project HybridNanoDev (Grant  No.RBFR1236VV), EU FP7/2007-2013 under REA grant agreement no 630925 -- COHEAT, MIUR-FIRB2013 -- Project Coca (Grant
No.~RBFR1379UX), and the COST Action MP1209.
G. D. thanks also CNR-SPIN via Seed Project PGESE003.


\begin{thebibliography}{77}
\bibitem{footnote-distance} A more general situation, however, does not affect our conclusions and physical results.
\bibitem{footnote-origin} Note that the tunneling properties do not depend on the choice of the origin of the coordinate system, but only on the relative distance between the QPCs.
\bibitem{footnote-phonons} Note that in this work we consider only the electronic contribution to thermal transport. We neglect phonon contributions since for the range of temperature considered these are negligible.
\bibitem{footnote-fermi} The Fermi distribution is here defined having the zero energy at the chemical potential
\end{thebibliography}
\end{document}